# Comparative Analysis of Segment Anything Model and U-Net for Breast Tumor Detection in Ultrasound and Mammography Images

Mohsen Ahmadi[1], Masoumeh Farhadi Nia[2], Sara Asgarian[3], Kasra Danesh[1], Elyas Irankhah[4], Ahmad Gholizadeh Lonbar[5], Abbas Sharifi[6,*]

*1-Department of Electrical Engineering and Computer Science, Florida Atlantic University, FL, USA*
*2-Department of Electrical and Computer Engineering, University of Massachusetts Lowell, Lowell, MA, USA*
*3- Cellular and Molecular Endocrine Research Center, Research Institute for Endocrine Sciences, Shahid Beheshti University of Medical Sciences, P.O. Box 19195-4763, Tehran, Iran*
*4-Department of Mechanical Engineering, University of Massachusetts Lowell, Lowell, MA, USA*
*5- Department of Civil, Architectural and Environmental Engineering, Missouri University of Science and Technology, Rolla, MO, USA*
*6- Department of Civil and Environmental Engineering, Florida International University, Miami, FL, USA*
*Corresponding author: asharifi@fiu.edu

**Abstract**

In this study, the main objective is to develop an algorithm capable of identifying and delineating tumor regions in breast ultrasound (BUS) and mammographic images. The technique employs two advanced deep learning architectures, namely U-Net and pretrained SAM, for tumor segmentation. The U-Net model is specifically designed for medical image segmentation and leverages its deep convolutional neural network framework to extract meaningful features from input images. On the other hand, the pretrained SAM architecture incorporates a mechanism to capture spatial dependencies and generate segmentation results. Evaluation is conducted on a diverse dataset containing annotated tumor regions in BUS and mammographic images, covering both benign and malignant tumors. This dataset enables a comprehensive assessment of the algorithm's performance across different tumor types. Results demonstrate that the U-Net model outperforms the pretrained SAM architecture in accurately identifying and segmenting tumor regions in both BUS and mammographic images. The U-Net exhibits superior performance in challenging cases involving irregular shapes, indistinct boundaries, and high tumor heterogeneity. In contrast, the pretrained SAM architecture exhibits limitations in accurately identifying tumor areas, particularly for malignant tumors and objects with weak boundaries or complex shapes. These findings highlight the importance of selecting appropriate deep learning architectures tailored for medical image segmentation. The U-Net model showcases its potential as a robust and accurate tool for tumor detection, while the pretrained SAM architecture suggests the need for further improvements to enhance segmentation performance.

## 1. Introduction

Breast cancer poses a challenge to global healthcare systems and remains the main reason for cancer-related fatalities in women in numerous countries worldwide. As per the World Health Organization (WHO), breast cancer is identified as either the primary or secondary cause of death among women in approximately 95% of countries surveyed [1]. Breast lumps is further split into two types: malignant cancer and benign (non-cancerous). Unlike malignant tumors, which form in or near the breast tissue and grow quickly, benign tumors have a slow growth rate and do not invade surrounding tissues or metastasize to distant parts of the body. Ductal and lobular carcinoma are the two predominant subtypes of breast cancer. In contrast to lobular carcinoma, that starts in the milk-producing glands and spreads to the nipple more slowly, ductal carcinoma begins in the breast ducts and spreads to the nipple [2]. Detecting breast cancer at an early stage holds paramount in reducing both the mortality rates associated with the disease and its overall impact on the population [3]. Breast cancer can be identified in its early stages through various means such as self-examination to monitor changes in breast shape, weight, and texture, clinical examination by a healthcare professional, and advanced techniques such as mammograms, 3D mammography, and ultrasound. These methods can detect breast cancer up to two years before it can be detected by touch. Mammography and ultrasound imaging are commonly employed imaging techniques for the detection of cancers [4]. Ultrasound is a valuable medical imaging technique that utilizes sound waves to identify and evaluate abnormalities in breast tissue. While it is commonly used as a supplementary tool alongside physical examinations and mammography, ultrasound serves as a reliable method for screening and detecting breast cancer. Unlike other imaging modalities, ultrasound does not employ ionizing radiation, making it a safer option. Its real-time visualization capability, affordability, and non-invasive nature contribute to its widespread usage in detecting breast masses. Furthermore, ultrasound imaging allows for the evaluation of breast density, equipping healthcare providers with supplementary data for precise diagnosis and effective treatment strategizing [5].

It is worth mentioning that according to [6], Ultrasound images are used in one out of every four studies, and the percentage is rising. [7] claims that ultrasound expands the scope of cancer detection by %17. Over the past two decades, there has been interest in the study of automatic breast ultrasound (BUS) image segmentation. Technologies can be categorized as fully automated or semi-automated based on user interaction requirements. Semi-automated approaches necessitate the specification of a region of interest (ROI) by the radiologist, such as the lesion, initial border, or a seed point within the lesion. In contrast, fully automated segmentation techniques operate without user input and often utilize prior knowledge of breast oncology and ultrasound as constraints for accurate segmentation. These advancements aim to enhance efficiency and accuracy in breast ultrasound image analysis [8]. Mammography is a type of imaging technique used for screening and diagnosing breast conditions, particularly breast cancer. It involves taking X-ray images of the breasts to detect any abnormalities, such as tumors or suspicious areas. Studies have shown that it contributes to decreasing cancer mortality and treatment-related morbidity. Both equipment and expertise are necessary for high-quality mammograms [9]. Both ultrasound and mammography imaging serve as modalities employed by radiologists to detect cancers and aid in their diagnosis. Expert clinicians must therefore manually visualize whole-slide images for the purpose of grade cancer, which is a laborious task and might additionally result in wrong decisions based on the various diagnostic criteria utilized globally. Similarly, the distribution of emphasis among different aspects during the comprehensive diagnosis often results in limited objectivity and reproducibility of assessment outcomes, primarily due to variations in doctors' technological capabilities and expertise. Consequently, the integration of an automated computer-aided decision (CAD) support system can effectively mitigate false positives and enhance diagnostic accuracy with reduced reliance on expert clinician interaction. The outcome appears to be an efficient approach for automatically detecting malignant tumors beyond the subjectivity of manual image analysis[10], [11]. CAD is a computerized system that provides support to healthcare practitioners in identifying illnesses, particularly cancer, by evaluating medical images such as MRIs, X-rays, CT scans, ultrasounds, and microscopic images. Autonomous CAD systems are being developed with the intention of accurately and quickly extracting a particular disease. To achieve this goal, four fundamental steps including preprocessing, segmentation, feature extraction [12], and classification [13,14] are implemented. Through the application of Deep Learning, there was a advancement in image processing including image segmentation. Artificial Neural Networks that are commonly utilized in Deep learning mostly is using Convolution Neural Network (CNN) architectures. Fully connected (FC) layers, and pooling are common building components used in the creation of CNNs. AlexNen, ResNet, and VGGNet are the three main CNN models. Another common neural-network-oriented technique utilized for medical image segmentation is U-NET, a two-path neural network for decoding and encoding.

## 2. Related works

The following Table 1 provides a comprehensive overview of various studies conducted on breast tumor segmentation in medical image analysis. Each study focuses on different datasets, techniques, and algorithms to obtain accurate tumor boundary delineation and classification. Methods such as watershed algorithm, fuzzy C-means algorithm, vector quantization, and region growth are employed for segmentation, while classification techniques include K-Nearest Neighbor (kNN), Support Vector Machine (SVM), and Artificial Neural Network (ANN). The results indicate varying accuracy rates ranging from 85.57% to 98%. Furthermore, advanced approaches like deep learning networks, modified U-Net architectures, and hybrid CNN-transformer networks have been proposed in recent years, showcasing promising results and outperforming previous methods. These studies collectively contribute to the advancement of breast tumor segmentation techniques, aiming to enhance the precision and effectiveness of diagnosis in medical imaging applications. It is worth mentioning that the utilized datasets in most of them is Mammographic Image Analysis Society (MIAS).

**Table 1**: A scope of reviewed studies in Breast cancer

| Author | Year | Dataset | Method | Result/ accuracy rate |
|---|---|---|---|---|
| Huang et al. [15] | 2006 | US Ultrasound Images Database | • Initially, the watershed algorithm was employed to acquire the initial active contour model, followed by the optimization of the energy function to precisely delineate the boundaries of the tumor. | saving time needed to sketch a precise contour with very high stability |
| Srivastava et al [16] | 2013 | MIAS | • The segmentation procedure employs a three-class fuzzy C-means algorithm.<br>• Techniques for classifying cancerous and non-cancerous cells included SVM, (kNN), and Artificial Neural Network (ANN). | Accuracy rate of 85.57% |
| Kekre et al [17] | 2013 | Ultrasound Images of Breast | • For the segmentation of breast images in ultrasound (US), a three-stage algorithm was employed. It began with the utilization of a clustering technique based on Vector Quantization (VQ) using the Linde Buzo Gray (LBG) algorithm, along with Kekre's | The consultant radiologist believes that KEVR produces the best results out of |

| | | | Proportionate Error (KPE) and Kekre's Error Vector Rotation (KEVR) codebook generation algorithms. This was followed by sequential cluster merging to achieve image segmentation. | all clustering techniques. |
|---|---|---|---|---|
| Vishrutha et al [18] | 2015 | Mini-MIAS | • A segmentation approach called region growth is utilized to recognize the ROI.<br>• The segmented images/regions were processed using the discrete wavelet transform approach to extract features. Finally, the mammography pictures were classified as malignant or benign utilizing the SVM classifier algorithm. | Accuracy rate of 92% |
| Pashoutan et al [19] | 2017 | MIAS | • Four different methods including Gabor wavelet transform, Wavelet transform, , Gray-Level Co-occurrence Matrix (GLCM, and Zernike moments)—were used to obtain features. | Accuracy rate of 94.18% |
| Hariraj et al [20] | 2018 | Mini-MIAS | • Fuzzy Multi-layer was employed at the pre-processing stage to remove wedges and labels from photos.<br>• K-means clustering was used for the segmentation step. | Accuracy rate of 98% |
| Eltoukhy et al [21] | 2018 | IRMA and MIAS | • The implementation involves extracting features using the curvelet transform and moment theory, employing a 10-fold cross-validation scheme. | 93.3% and 90.6% accuracy for the IRMA and MIAS databases |
| Sarosa et al [22] | 2019 | MIAS | • The classification methods of GLCM and Backpropagation Neural Network (BPNN) were examined.<br>• Features were extracted from pre-processed pictures using GLCM.<br>• To identify whether or not the input image is normal, BPNN was utilized. | Accuracy rate of 90%. |
| Arafa et al. [23] | 2019 | MIAS | • The ROI was determined using the Gaussian Mixture Model (GMM).<br>• The segmented ROI was classified into normal, abnormal, benign, and malignant using the SVM classifier technique. | Accuracy rate of 90% |
| Yang et al [24] | 2019 | IVUS (Intravascular Ultrasound) dataset | • A 3D cardiovascular image segmentation framework, known as Dual-Path U-Net (DPU-Net) architecture, was utilized. | Accuracy rate of 92.5% |
| Hu et al [25] | 2019 | BUS Images from Department of Ultrasound, Fudan University Shanghai Cancer Center, China | • Deep Fully Convolutional Network (DFCN) and Phase-Based Active Contour (PBAC) methods are utilized | For 40 MHz the media segmentation accuracy increased from 0.775 to 0.863 and for 20 MHz increased from 0.855 to 0.921. |
| Zebari et al [26] | 2019 | images in a one dataset of 200 monograph cases (60 normal cases | • Wavelet transforms<br>• Image enhancement for segmentation and feature extraction were employed. | Segmentation: Ac = 90.5 |

| Author | Year | Dataset | Method | Result |
|---|---|---|---|---|
| | | and 140 abnormal cases) | | Feature extraction: PSNR = 69.95 |
| Farhan and Kamil [27] | 2020 | Mini-MIAS | • In addition, the local binary pattern (LBP) and the histogram of oriented gradient (GLCM) approaches have been utilized to obtain features.<br>• In the final step, cancerous and non-cancerous cells were classified using SVM and kNN classifier algorithms. | DFCN 68.8%<br><br>DFCN+PBAC 71.9% |
| Eltrass and Salama [28] | 2020 | MIAS | • Features have been extracted using a contourlet transform technique based on wavelets.<br>• SVM classifier algorithm was employed | 90.3% |
| Saeed et al. [29] | 2020 | MAIS | • The ROI was segmented using a hybrid bounding box and region growing method.<br>• Statistical features including mean, standard deviation, skewness, and kurtosis as well as texture features like LBP and GLCM were both extracted as part of the features extraction process | 98.16% |
| Li et al [30] | 2020 | LiTS and CHAOS | • The proposed solution introduces the Attention-based Nested U-Net (ANU-Net) | accuracy of 95.45% was obtained for the first level and 97.26% for the second level |
| Mischi et al. [31] | 2020 | XPIE dataset | • Modified U-Net architecture is implemented<br>• (Ref: RRCNet: Refinement residual convolutional network for breast ultrasound images segmentation) | LiTS Liver 0.9815<br>CHAOS Spleen 0.9519<br>CHAOS Kindey 0.9400<br>CHAOS Liver 0.9423 |
| Rampun et al [32] | 2020 | (MIAS) and InBreast datasets | • The methodology integrates active contour, restricted contour growing with edge information, median filter, breast region segmentation with noise reduction, and investigates encoding techniques in binary-based local patterns for mammographic breast density classification | PT 10 |
| Shen et al [33] | 2020 | MIAS | • Breast Cancer diagnosis via Type-2 Fuzzy Learning hierarchical Fused Model with Deep Learning | PT 10 |
| Christopher et al. [34] | 2020 | MIAS and InBreast datasets | • The technique employed is Mammogram Enhancement using Nonlinear Unsharp Masking and L0 Gradient Minimization | EME = 3.89,<br>AME = 23.92,<br>SDME = 49.36 |
| Suganthi et al. [35] | 2020 | MIAS | • The method makes use of active contour and intensity-based thresholding techniques to identify the boundary separating the pectoral muscle region from the remaining breast area. Improved deep Res U-Net for pectoral muscle segmentation is used. | accuracy of 92.55% |
| Chen, Wang, and Chen [36] | 2020 | IN breast and CBIS-DDSM | • A mass segmentation approach based on an enhanced U-Net architecture is employed, which incorporates a multi-scale adversarial network. | dice of 81.64% for INbreast and 82.16% for CBIS-DDSM |
| Alzahrani. Y and Boufama [37] | 2021 | BUSI and UDIAT public databases | • For BUS image segmentation, DL method using a modified U-Net architecture is utilized | 98.44% |
| Civit-Masot J. et al. [38] | 2021 | DRISHTI-GS RIM-ONE-v3 | • DL for Medical Image Segmentation is used.<br>• Hardware Acceleration on SBCs (Google's Edge TPU) | Scenario 1:<br>BUSI: 0.995<br>UDIAT: 0.949<br>Scenario 2:<br>BUSI: 0.977 |

| | | | | UDIAT: 0.977 |
|---|---|---|---|---|
| Xue et al [39]. | 2021 | BUSI in Al-Dhabyani et al., 2020 + Collected Dataset | • Global Guidance Block (GGB) and Breast Lesion Boundary Detection (BD) Module are utilized. | Their network performs better than other approaches for segmenting medical images. |
| Ahmed et al [40] | 2023 | BUS dataset, RIDER Breast dataset, and CBIS-DDSM Mammogram dataset | • The BTS-ST network is a solution for Swin-Transformer (ST)-desiring breast tumor segmentation and classification.<br>• The fundamental architecture of BTS-ST incorporates a dual encoder that utilizes U-Net and Swin-Transformer, along with modules such as Spatial Interaction Block (SIB), Feature Compression Block (FCB), and Relationship Aggregation Block (RAB). | The suggested method obtained F1(0.908), F1(0.833), and F1(0.771) on the Private BUS dataset, RIDER Breast dataset, and CBIS-DDSM Mammogram dataset. |
| Qiqi et al [41] | 2023 | BUSI dataset, BUS dataset and Dataset B | • A combined network consisting of a CNN and a transformer is utilized for the segmentation of breast ultrasound images. | HCTNet achieves better performance against other segmentation networks on all three datasets. |

## 3. Methods and Materials
### *3.1. Medical image processing*

Medical image processing algorithms can help in the diagnosis and categorization of various breast abnormalities, such as masses and microcalcifications, by looking for patterns and features in images obtained from mammography, ultrasound, MRI, and CT scans. Support vector machines and artificial neural networks can be used to automate the detection and classification of breast lesions, improving the accuracy and consistency of diagnoses [42]. Technically, large databases of medical images can be analyzed by medical image processing algorithms to identify high-risk people who require additional testing and evaluation. This can help in breast cancer screening. Several risk factors, such as age, family history, and genetic changes, can be examined using machine learning and deep learning algorithms to predict the possibility of developing breast cancer [43]. Medical image processing can also be used to enhance the efficiency and accuracy of image segmentation, an important step in the interpretation of medical images. Finding and separating areas of interest from surrounding tissue, such as breast lesions, is the process of image segmentation. It should be emphasized that automated image segmentation utilizing machine learning can produce faster and more accurate diagnoses and can reduce the time and effort required for manual image segmentation.

### *3.2. Breast ultrasound images segmentation*

For the most part, automated image segmentation using algorithms and machine learning techniques can help to improve the consistency and accuracy of breast ultrasound image analysis, whereas manual image segmentation can be time-consuming and subject to inter- and intra-observer variability, leading to potential errors and inconsistencies in the analysis.

There are various benefits to this strategy over manual segmentation. Firstly, it can reduce the time and effort required for image analysis, particularly for large datasets. Second, it may boost the accuracy and consistency of the analysis since it is less susceptible to biases and mistakes brought on by human interpretation. Finally, it can help in the detection of minute or subtle irregularities that manual segmentation may have overlooked [44]. Consequently, many techniques have been created for segmenting breast ultrasound images, including thresholding, edge detection, region-based techniques, and machine learning-based techniques. Regions of interest are isolated from the surrounding tissue based on pixel intensities by changing a threshold value. Finding edges in a picture and separating portions along those edges is the technique of edge detection. Using area-based algorithms, the image is separated into regions, and each region is then classified as having or not having a lesion depending on several variables. Using a massive dataset of annotated ultrasonic images, machine learning-based techniques instruct a computer algorithm to discern the features and patterns that distinguish different tissue types and lesions. After training, the algorithm can automatically distinguish between the various breast ultrasound image regions [45]. Despite these developments, adopting automated image segmentation for breast ultrasound images still has its drawbacks. For instance, thresholding approaches frequently depend on the overall quality of the image and may not be effective with photographs that have poor contrast or uneven illumination. Similarly, when boundaries between structures are ambiguous or poorly defined, edge detection may be unable to precisely separate regions. Despite often being more resistant to fluctuations in image quality, region-based approaches may have trouble with intricate anatomical features or overlapping tissues. Although very promising, machine learning-based solutions are not without their own set of difficulties.

Large, annotated datasets are necessary for training a machine learning model for this activity, but their acquisition can be time- and money-consuming. Additionally, these models might overfit the training set, which would make them less generalizable to brand-new, untested images. Additionally, these algorithms frequently take the form of "Black Boxes," making it challenging to understand their decision-making process, which is crucial in the therapeutic setting. Notwithstanding these difficulties, it is nevertheless important to conduct ongoing research and development in this area since the advantages of automated picture segmentation, such as enhanced effectiveness and accuracy, have the potential to enhance patient care. There are solutions being investigated to get over these restrictions, such as integrating several segmentation techniques or making machine learning models easier to understand. Furthermore, adopting more sophisticated machine learning methods like deep learning may lead to improved performance. The classification of lesions or the automated report creation are two more areas where future research may use automated image segmentation. Additionally, incorporating machine learning into the healthcare sector has the potential to lead to a paradigm change that would improve patient outcomes and streamline the delivery of healthcare. For a condition like breast cancer, it might enable early identification and intervention, lowering its morbidity and mortality. Even with the present limitations, the promise held by automated image segmentation and machine learning in improving healthcare can't be underestimated, and further research in this field is anticipated to yield advancements.

Basically, the differentiations discernible in breast ultrasound images associated with malignancies versus benign conditions are vast and nuanced. These differences manifest in various aspects, including morphology, echogenicity, edge characteristics, posterior features, and associated findings, each bearing implications for the subsequent diagnostic process and patient care (see Figure 1).

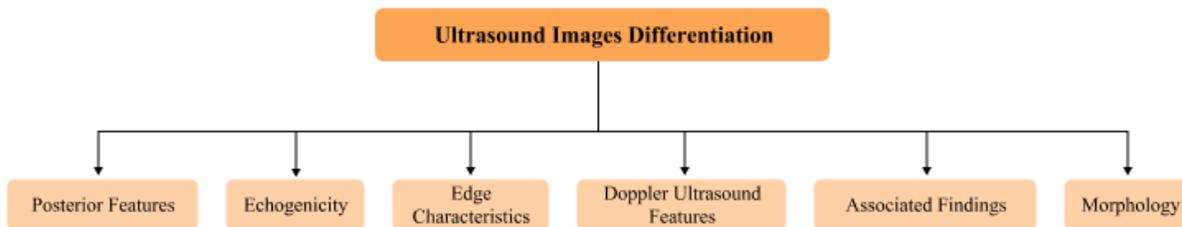

**Figure 1**: A block diagram of ultrasound images differentiation

Although there is some overlap between the ultrasound findings of benign and malignant tumors, these traits are not absolute. In order to provide a precise diagnosis, breast ultrasound results are often combined with patient history, physical examination, mammography, and, if necessary, more invasive procedures like biopsy. In addition, advances in breast cancer lesion definition and diagnostic precision are being made thanks to developments in ultrasound technologies like elastography and contrast-enhanced ultrasonography.

### *3.3. The CSAW-S Dataset*

The CSAW-S dataset, which is restricted access and accessible at [69] has been painstakingly curated and consists of mammography images that have been annotated by both experts and non-experts, focusing on breast cancer and breast anatomy. This dataset can be used for a variety of tasks, including the replication of our study, repurposing for different semantic segmentation tasks, or combining with other datasets on breast cancer, like DDSM and INbreast, to build up a large collection of mammograms with tumor annotations.

### *3.4. Breast Tumor Ultrasound Dataset*

In this study, we used a dataset [70] comprised of ultrasound images of breasts obtained from 600 female subjects aged between 25 and 75. The dataset consists of 780 PNG images, each having a resolution of 500 x 500 pixels, classified into three classes: normal, benign, and malignant. Specifically, there are 133 normal images, 487 benign images, and 210 malignant images.

### *3.5. Data Collection*

The CSAW-S collection contains images from 172 distinct mammography screenings for breast cancer. It is divided into two sets: a test set, made up of 26 images from 23 different patients, and a training/validation set, made up of 312 pictures from 150 patients. The distribution of the classes between the training and test divides is ensured by the data division. The dataset includes 338 high-definition grayscale images of the skull taken from the MLO (Mediolateral-Oblique) and CC (Cranial-Caudal) viewpoints. These screening images were chosen at random from the CSAW database, a repository of screenings conducted with Hologic technology in Stockholm between 2008 and 2015. The dataset has annotations for 13 classes, including a background class, ten classes that reflect different aspects of breast anatomy, and a class for cancer (relating to expert annotations). The cancer annotations were created by three radiological professionals. In order to accomplish the goals of semantic segmentation (identifying anatomy for quality control and analyzing potential cancer correlations), discussions with these experts helped us determine the classes for annotation. Non-specialists annotated the supplemental lessons in the training package. Both the annotations for the cancer class and the supplemental classes for the test set were handled by the experts. Notably, the experts did not annotate the training sets complementary classes. Despite never having seen mammography images before, the novices participated in a quick training session. Notable imbalances can be seen in the supplemental courses, with some classes only seldom showing up.

### *3.6. Performance metric*

A machine learning model or algorithm's performance is assessed using performance metrics. The effectiveness of the model's classification or outcome prediction based on the input data is measured using these metrics. Performance measures are frequently used to assess the precision and efficacy of medical image processing algorithms in the context of breast cancer diagnosis and screening [46] [47]. Predominantly, there are several performance metrics used in machine learning, including accuracy, precision, sensitivity, recall, and F1_measure. These metrics can be calculated using the confusion matrix, which is a table that compares the actual outcomes with the predicted outcomes [48] [49] .

### *3.7. loss function*

Loss functions are crucial in the field of machine learning because they direct the optimization of a model's parameters. In this work, an attempt was made to select a loss function that would accurately classify and segment the photos [50] [51]. Many versions of these functions are employed based on the specific problem and method at hand. In medical image processing for breast cancer, loss functions such mean squared error (MSE), dice loss, binary cross-entropy, and categorical cross-entropy are widely utilized. Binary cross-entropy has been employed for binary classification problems, such as distinguishing between benign and malignant breast cancers. It determines the difference between each class's actual labels and expected probabilities. Categorical cross-entropy is utilized for multi-class classification issues such as recognizing different types of breast lesions[52]. Dice loss is additionally frequently applied to image segmentation issues, such as separating breast tumors from surrounding tissue in medical photographs. It calculates how much of the predicted segmentation mask overlaps with the actual segmentation mask. When dealing with regression issues like estimating the size or volume of breast lesions, mean squared error, or MSE, is frequently used.

We have considered the possibility that the selection of a suitable loss function may have a substantial effect on the performance of a machine learning system. To reduce false positives, for instance, a loss function that emphasizes specificity over sensitivity may be applied. Alternatively, if detecting all probable cancer cases is the top priority, sensitivity over specificity may be chosen as the loss function of choice. A machine learning algorithm for medical image processing in breast cancer detection and screening may perform better or worse depending on a variety of factors in addition to the loss function that is selected. These include the algorithm design, image quality, and the imaging modality selection [53] [54].

### 3.8. CNN training

In order to lessen the gap between expected outcomes and training data outputs, CNN training technically comprises gradually altering the network's parameters. For this, a stochastic gradient descent optimization technique is typically used, which adjusts the network's weights and biases in accordance with the gradient of the loss function with respect to these parameters. The training process involves a number of important components. The training data must first go through preprocessing to ensure that it is properly cleaned and normalized for input into the network. After that, the network architecture must be built and given the correct initial biases and weights. The optimization algorithm then modifies the weights and biases of the network as it is trained using the training data to minimize the loss function. The model is subjected to this process over a number of epochs until it converges or meets a predetermined stopping threshold. [55].Several methods can be utilized during training to enhance network performance and avoid overfitting. These include regularization approaches that help prevent the network from forgetting the training data as well as improve generalization to new input, such as dropout, batch normalization, and data augmentation. A network's performance can also be optimized by adjusting hyperparameters like learning rate, batch size, and number of layers. [56] [57]. One of the challenges of CNN training in this work, which can be seen in the medical image processing for breast cancer, was the limited availability of annotated data. To circumvent this issue, transfer learning, a technique for improving a previously trained network on a smaller dataset, can be utilized to boost the network's performance. Additionally, generative adversarial networks (GANs) can be used to create artificial training data, which can help increase the variety of the training data and the generalization of the network. [58].

### 3.9. Pre-training models

Convolutional neural networks (CNN), a kind of deep learning algorithm that can recognize intricate patterns in images, are frequently used in the pre-training phase. Large datasets of breast imaging data that have been classified as either malignant or non-cancerous are used to train these networks. Through this procedure, the CNN can learn to identify patterns that are suggestive of malignancy, such as asymmetrical boundaries, abnormal forms, and grouped microcalcifications [59] [60]. The trained model is adjusted using a smaller dataset of breast images that have been annotated by medical experts once the pre-training procedure is finished. In order to improve CNN's ability to detect certain patterns in the labeled images, this fine-tuning procedure entails changing the CNN's weights. The outcome is a very precise model that can be used to identify breast cancer in fresh images. Data augmentation, which generates new images by making changes to existing photos, might improve the accuracy of pre-training models for the diagnosis of breast cancer by combining the predictions of many models. These techniques help reduce the potential for overfitting and increase the model's overall accuracy [61].

## 3.10. U-Net model

The U-Net deep learning architecture has shown promise in successfully segmenting and categorizing medical images, particularly in the diagnosis of breast cancer. It is made up of a gap between an encoder and a decoder. While the encoder portion of the network is identical to a typical CNN, the decoder section aims to reconstruct the original image and provide a segmentation mask that identifies the regions of interest in the image. The U-Net takes advantage of skip connections to maintain spatial information and give more accurate segmentation masks, allowing the decoder to access information from earlier layers of the encoder [62]. The task of reconstructing the original image and producing a segmentation mask that identifies the image's regions of interest is performed by the network's decoder component. The segmentation mask can be used to discern between cancerous and healthy tissue in medical images. One of the U-Net's distinctive features is the use of skip connections between the encoder and decoder components. This preserves spatial information and improves the accuracy of the segmentation mask by allowing the decoder to access data from earlier encoder layers. The U-Net is optimized during training by using a loss function that determines the difference between the expected segmentation mask and the actual segmentation mask. The U-Net can learn on a small amount of data via data augmentation techniques like flipping, rotating, and scaling the input images, which expand the size of the training dataset.

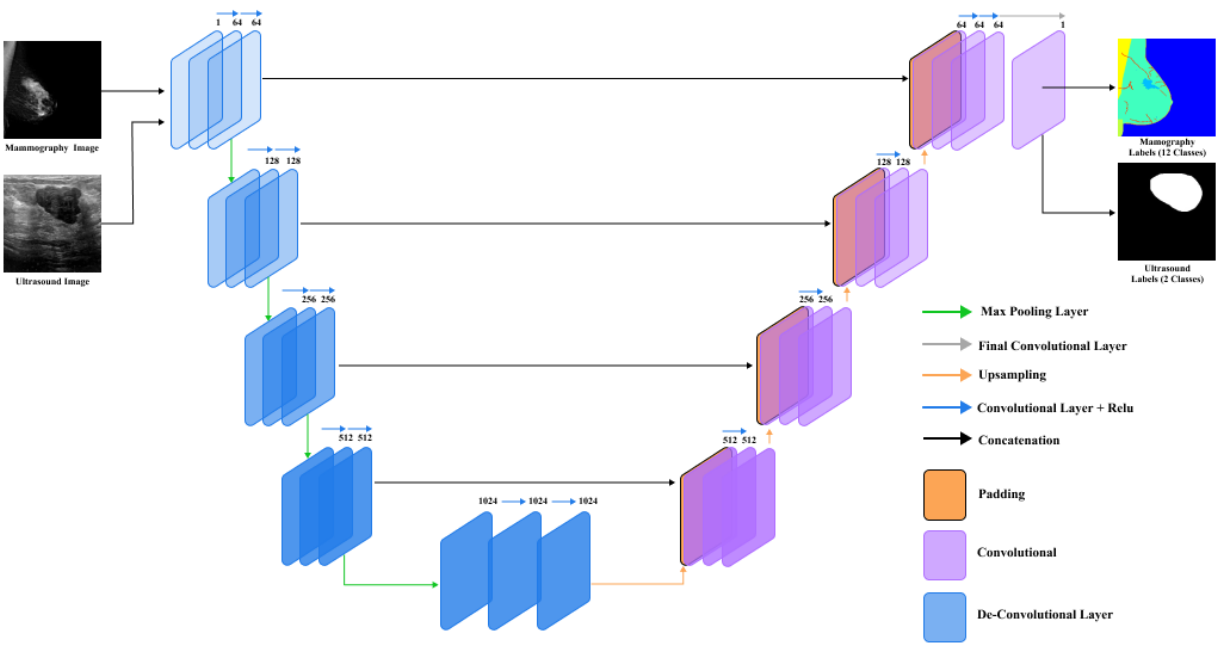

**Figure 2:** U-Net Architecture

The U-Net is also more efficient in terms of processing resources since it employs fewer parameters than other deep learning systems. With regard to medical image analysis, the U-Net is a promising tool, notably for the early identification and treatment of breast cancer [63-66].

### 3.11. Segment Anything Model (SAM)

SAM is essentially a deep learning architecture designed for the semantic segmentation of images, with a particular emphasis on segmenting medical images [67]. Its foundation is an encoder-decoder network with skip links, much like the SegNet architecture. The encoder half of the network reduces the spatial resolution of the input image by applying convolutional and pooling layers, while the decoder portion of the network reconstructs the original image from the feature maps produced by the encoder using up sampling and deconvolutional layers. Due to the skip connections between the relevant layers of the encoder and decoder during the down sampling process, the spatial information of the original image is preserved. Fundamentally, the Segment Anything Model is a comprehensive framework composed of many components, including a number of pre-processing tools, a substantial library of pre-trained models, and an understandable API enabling straightforward connection with other software systems. The pre-processing tools also include a variety of picture enhancement and noise reduction methodologies, as well as image normalization and registration tools, to preserve uniformity across diverse imaging modalities. The segmentation of organs, tumors, and lesions in various imaging modalities like CT, MRI, and ultrasound are just a few of the medical imaging applications that make use of the pre-trained models. The Model's functionality is mostly focused on image segmentation, which is the act of dividing an image into useable segments or regions, each of which corresponds to a specific object or component of an object. The model achieves this by employing advanced deep learning methods, such convolutional neural networks, to analyze and classify different portions of the image. The Segment Anything Model can precisely identify and classify different structures inside medical images, making it a useful tool for improving diagnostic accuracy and treatment planning in a variety of medical professions. Also, it has a wide range of uses in medical segmentation and has advanced this discipline. Medical segmentation is a crucial stage in the interpretation of medical images due to how it enables doctors to separate and recognize various structures, such as organs, tumors, and lesions, within the images. For a correct diagnosis, treatment planning, and condition monitoring, accurate segmentation of medical images is crucial. The segmentation of brain tumors from MRI images is one of the model's main uses. This is a crucial stage in the diagnosis and treatment of brain tumors because precise localization and measurement of the tumor's size are necessary for effective treatment planning. It has been demonstrated that the Segment Anything Model produces reliable segmentation findings that are correct, assisting physicians in deliberations about patient treatment. Likewise, the segmentation of the heart and blood arteries from cardiac CT scans, which is crucial for the diagnosis and treatment of cardiovascular disease, has been accomplished with the Segment Anything Model. Clinicians can spot anomalies and gauge a patient's risk of suffering problems like a heart attack or stroke by precisely segmenting the heart and blood arteries. Clinicians may establish accurate diagnoses and create effective treatment programs with the help of the Segment Anything Model, which has been demonstrated to deliver precise segmentation results.

Meanwhile, one of the Segment Anything Model's primary contributions is its capacity to boost the effectiveness and precision of medical segmentation. Medical segmentation has always required human annotation of big datasets, which is a time-consuming and labor-intensive operation. Clinicians can do segmentation jobs fast and accurately, even with little training data, according to the Segment Anything Model's pre-trained models and user-friendly API. Since medical professionals can base their judgments on correct segmentation results, this has important implications for enhancing patient care and outcomes [68]. On a technical basis, the Segment Anything Model has shown adept at segmenting a variety of medical imaging domains, including brain tumors and cardiovascular structures, due to its sophisticated deep learning capabilities. For better diagnosis and treatment of illnesses like breast cancer, it has been applied to breast imaging to delineate lesions in ultrasound and mammography images. Patient outcomes can be improved through the proper diagnosis of these lesions to direct biopsy procedures and surgical planning. Additionally, the model has proven useful in the field of orthopedics for segmenting bone and joint structures in CT and MRI scans. This has proven particularly beneficial in planning orthopedic surgeries and prosthesis fittings, improving both the accuracy and safety of these procedures. The model has even demonstrated potential in dental imaging, accurately distinguishing between various structures such as teeth, gums, and different types of dental implants. Along with these clinical uses, the Segment Anything Model also has extraordinary promise for use in academic settings. In the context of pathology and cell biology, it has been used, for example, to analyze microscopic images, advancing studies in fields like cancer biology and genetic abnormalities. It is also used in neuroimaging research to precisely segment different brain areas, improving our comprehension of the functional and structural organization of the brain. The model's capacity to adapt to different imaging modalities, anatomical structures, and even to microscopic images has made it an incredibly versatile tool in the field of medical imaging. Despite its considerable success, the continuous development of the Segment Anything Model is crucial, aiming to improve its generalizability across different patient populations, rare conditions, and less common imaging modalities. The future iterations of the model might also integrate more advanced deep learning techniques, like transformer models, to further enhance its performance. Additionally, the developers may focus on improving the model's interpretability, a critical aspect in the clinical context, making it easier to understand the reasoning behind its segmentation decisions. These potential enhancements, backed by continued research and clinical validation, can ensure that the Segment Anything Model remains a powerful tool for medical segmentation, contributing to both patient care and medical research.

## 4. Results

In this study, we conducted a comprehensive evaluation of two methods, U-Net and pretrained SAM, for tumor detection in breast cancer using two distinct datasets: Breast Ultrasound (BUS) images and mammography images. Our aim was to compare the performance of these methods and provide a detailed explanation of our findings (see Figures 3 and 4).

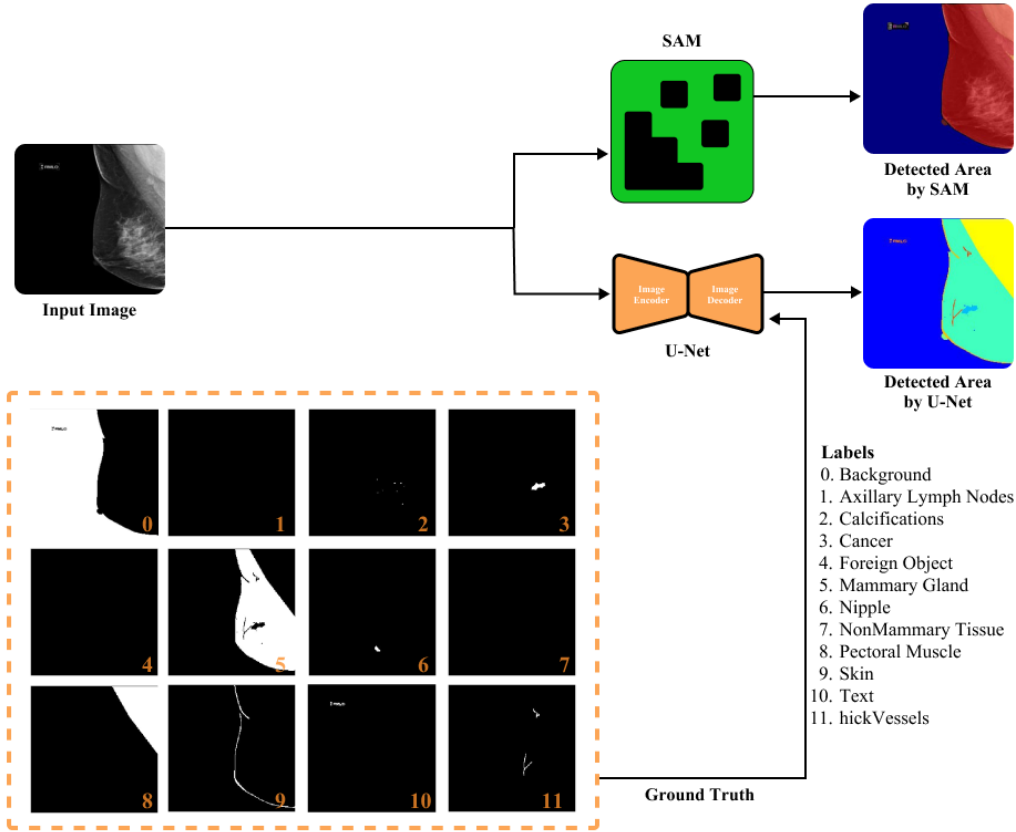

**Figure 3:** The detection process for mammography with 12 class image using SAM and U-Net

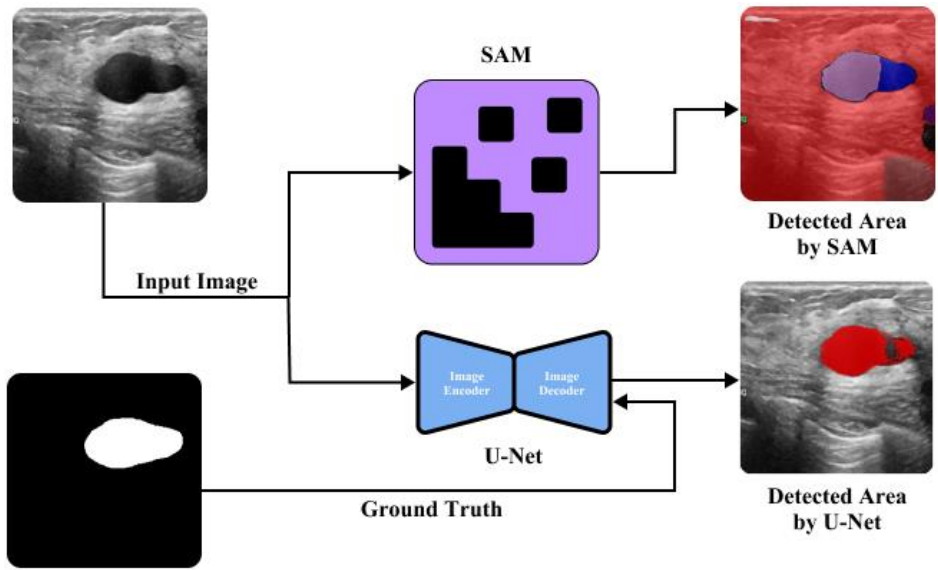

**Figure 4:** The detection process for Ultrasound with binary class image using SAM and U-Net

To achieve this, we meticulously carried out the detection process using both SAM and U-Net on the mammography dataset, which consisted of 12 class images. Figure 3 visually depicts the detection process for mammography, illustrating the intricate workings of SAM and U-Net in identifying tumors within the images. Additionally, we performed the detection process on the BUS dataset, which featured binary class images acquired through ultrasound imaging. The visualization of this process for ultrasound can be observed in Figure 4. By examining the outcomes of these evaluations, we aimed to gain valuable insights into the effectiveness and suitability of U-Net and pretrained SAM for tumor detection in breast cancer.

### *4.1. Results of segmentation using U-Net architecture with Ultrasound images*

For the U-Net method, we trained the model separately on the dataset using 300 epochs, a learning rate of 1e-4, and a cross-validation frequency of 10. The dataset was divided into 80% for training and 20% for testing purposes. Figure 5 illustrates the results of segmentation using the U-Net architecture on BUS images, including the original images and the detected tumor areas for three classes: normal tissue, benign tissue, and malignant tissue. The U-Net demonstrated its ability to successfully predict tumor areas in both benign and malignant tissues, utilizing pixel classifications. However, it is worth noting that post-processing is necessary to accurately separate the exact tumor area from the dispersed points.

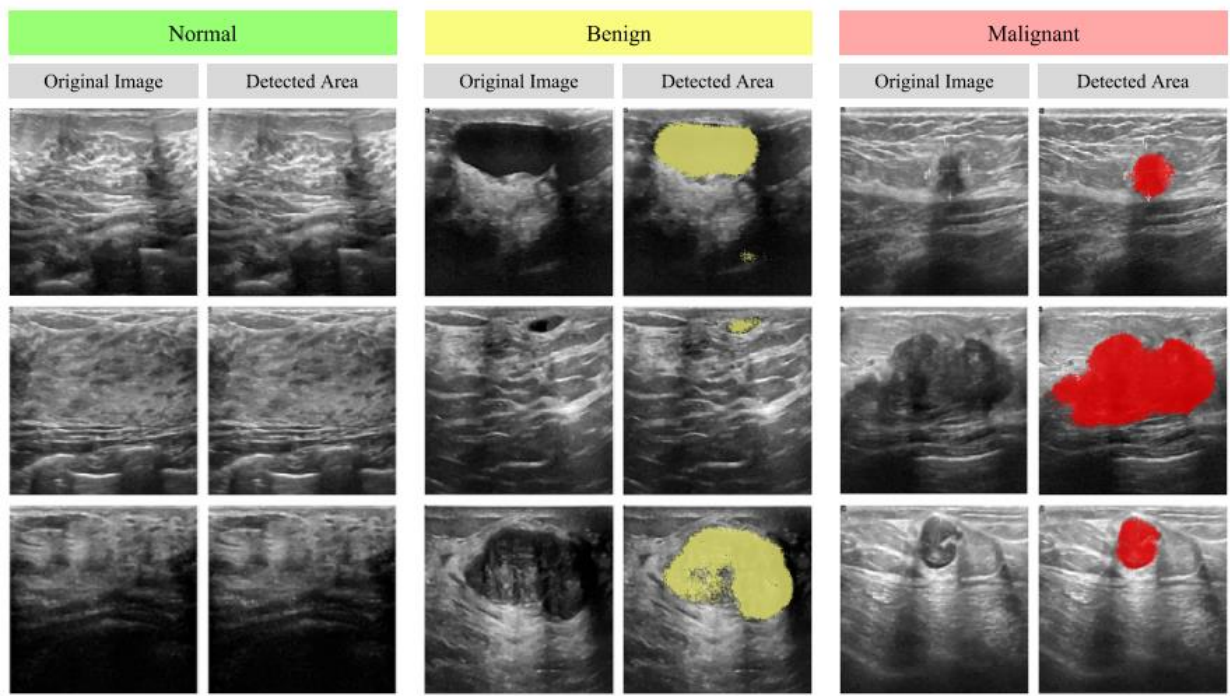

**Figure 5:** Results of segmentation using UNet method with Ultrasound images

To further evaluate the performance of the U-Net method, we employed ROC curves for the benign and malignant subsets, as depicted in Figure 6. ROC curves provide valuable insights into the trade-off between false positive rate and true positive rate, allowing us to gauge the effectiveness of tumor detection (positive class). Moreover, Figure 7 presents the corresponding area under the curve (AUC) values for the ROC curves.

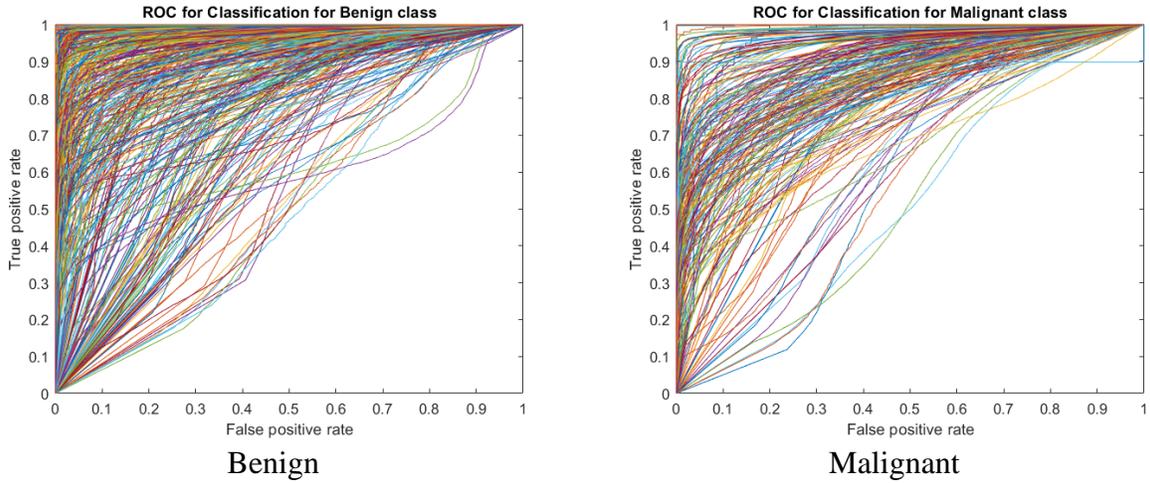

**Figure 6:** ROC curve for both benign and Malignant classes based on UNet using Ultrasound images

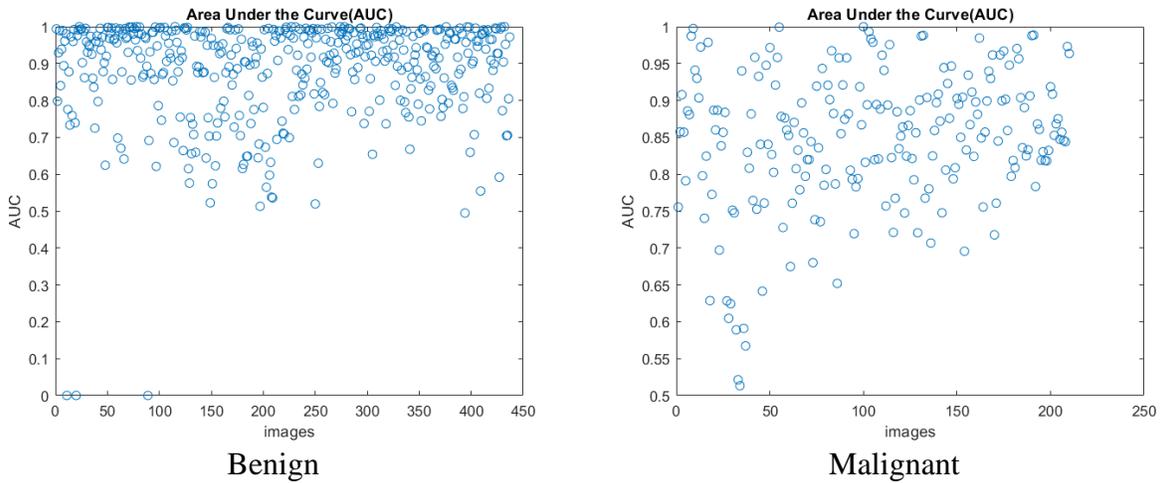

**Figure 7:** The Values of AUC for benign and Malignant classes based on UNet using Ultrasound images

The average AUC value for the benign class was recorded at 92.62%, indicating a high level of performance in identifying benign tumors. In the case of malignant tumors, the average AUC value slightly dropped to 84.82%, suggesting a relatively lower performance compared to benign tumors.

### *4.2. Results of segmentation using pretrained SAM architecture with Ultrasound images*

Moving on to the pretrained SAM architecture, we analyzed its results as shown in Figure 8. This figure showcases the detected areas and original images for the normal, benign, and malignant tumor classes. SAM exhibited the ability to separate images into different regions with distinct colors, enabling the identification of tumor areas in both benign and malignant tissues. However, it is important to consider the limitations of SAM. In the case of normal tissue, SAM tends to segment regions based on color intensity differences, which may not accurately correspond to actual tumor areas. SAM is also sensitive to text and notes present on the images, falsely considering them as tumor areas. Additionally, in many ultrasound images, SAM fails to identify any regions for separation, indicating limitations in its performance.

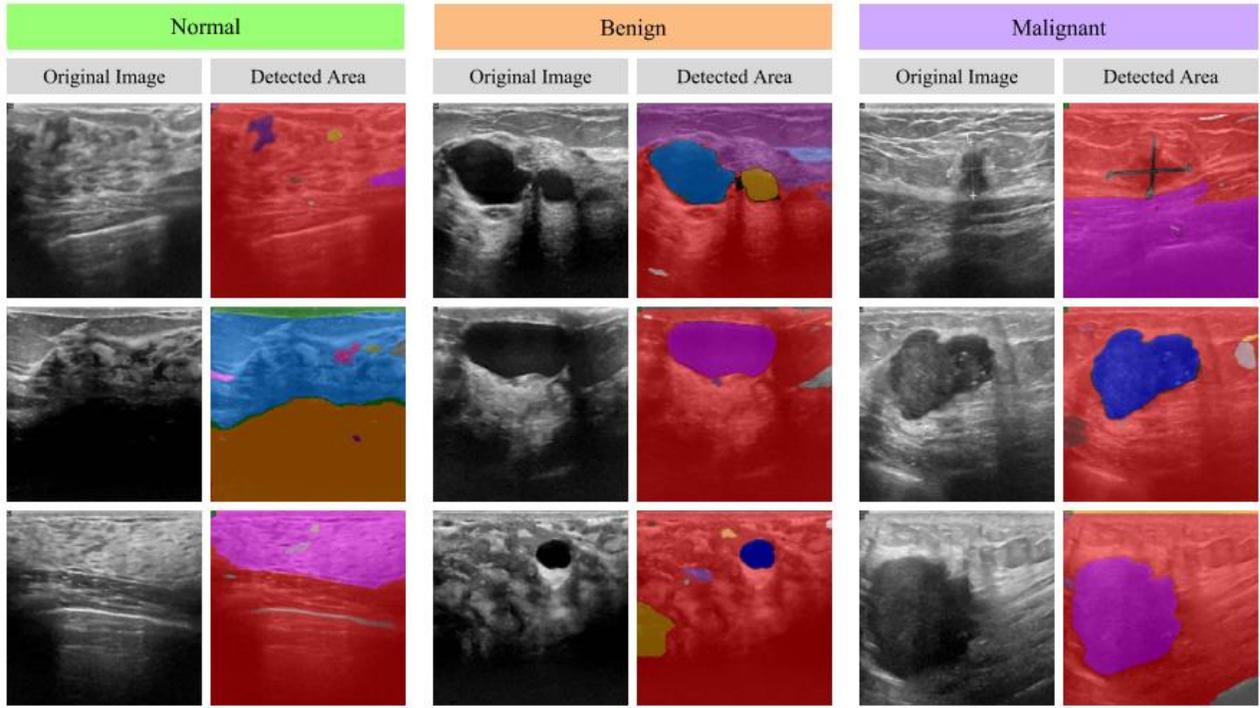

**Figure 8:** Results of segmentation using SAM method with Ultrasound images

### *4.3. Results of segmentation using UNet architecture with mammographic images*

Shifting our focus to the U-Net architecture applied to mammographic images, Figure 9 showcases the segmentation results for 12 classes, encompassing background, axillary lymph nodes, calcifications, cancer, foreign objects, mammary gland, nipple, non-mammary tissue, pectoral muscle, skin, text or notes on the image, and thick vessels. We used data augmentation method to increase number of images are prevent overfitting. The U-Net was trained in original mammography images, with categorical output images containing the 12 classes. The results highlight that the U-Net can successfully identify the tumor area when the image does not include the pectoral muscle. However, in images with the pectoral muscle, the tumor area is misclassified with the muscle due to color similarity. Additionally, the U-Net faces challenges in identifying thick vessels in the breast tissue, often detecting them with a limited number of points.

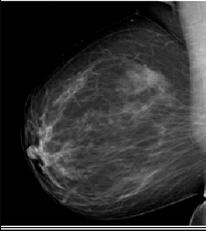

**Figure 9:** Results of segmentation using UNet method with mamography images

To provide a comprehensive assessment of the U-Net results, we generated ROC curves for all the classes, as depicted in Figure 10, and recorded the corresponding AUC values in Table 2. These metrics offer valuable insights into the U-net's performance across different classes, aiding in the evaluation and comparison of its effectiveness in tumor detection. Finally, Figure 11 presents the results of SAM on mammographic images. The SAM network demonstrates its capability to segment tumor tissue into several regions. However, it shares some limitations observed in the BUS images, such as sensitivity to text and notes on the images and the inability to identify regions for separation in certain cases.

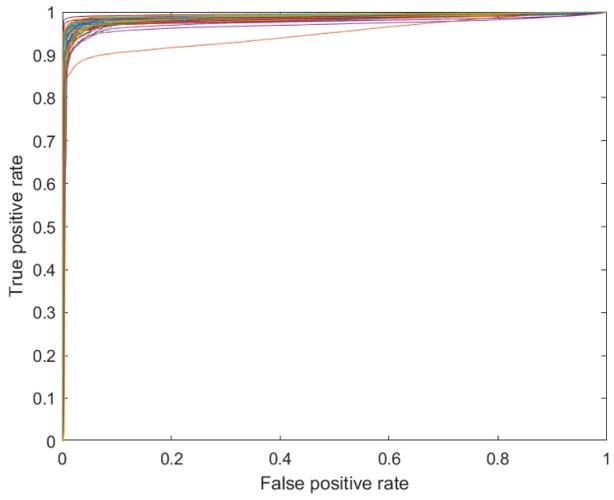
background

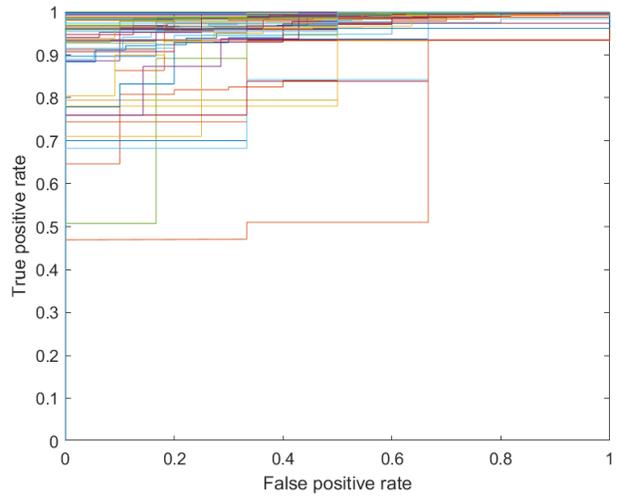
Axillary Lymph Nodes

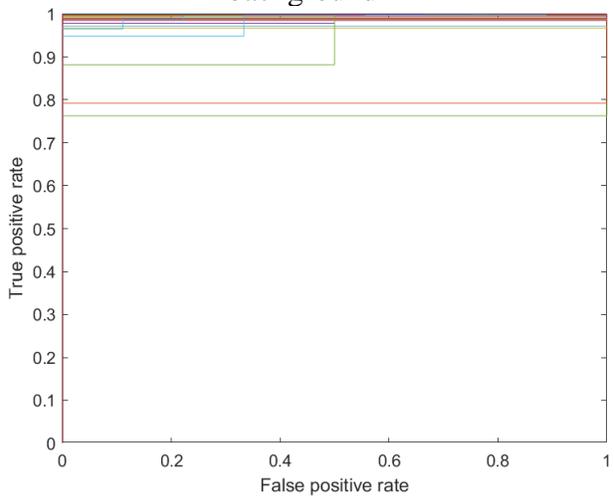
calcifications

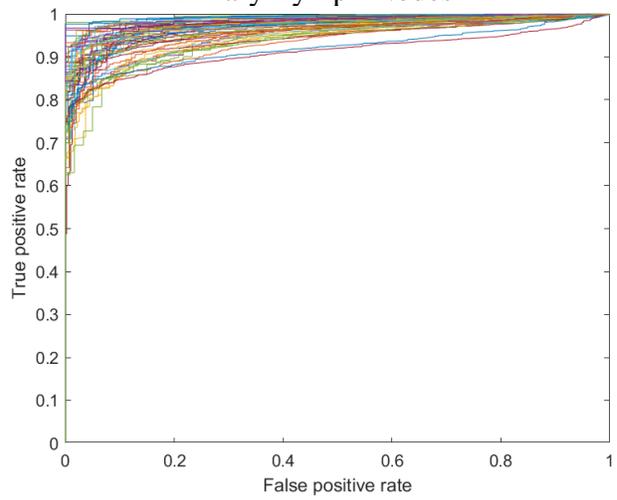
Cancer

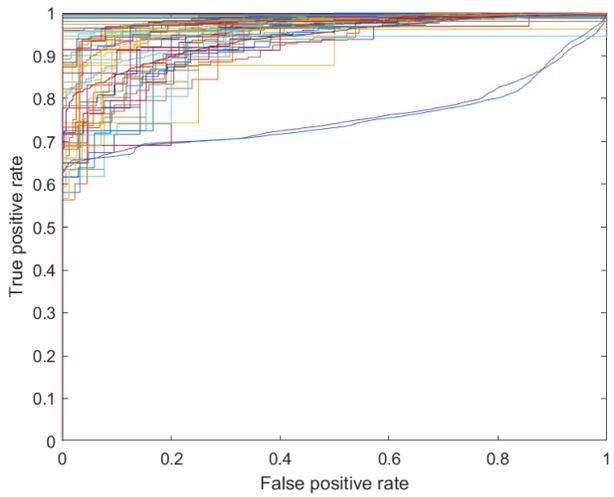
foreign Object

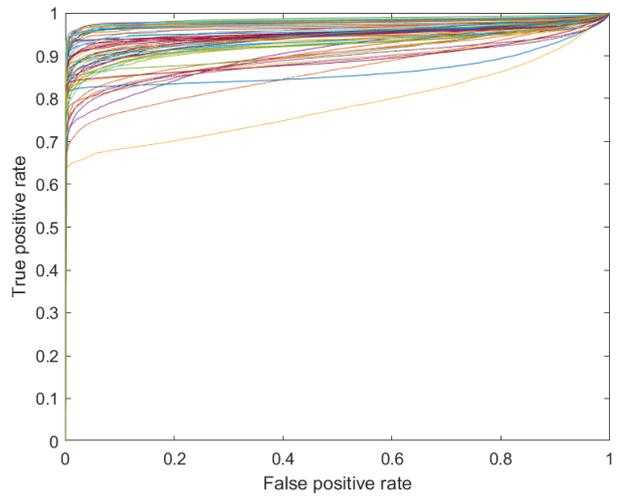
mammary Gland

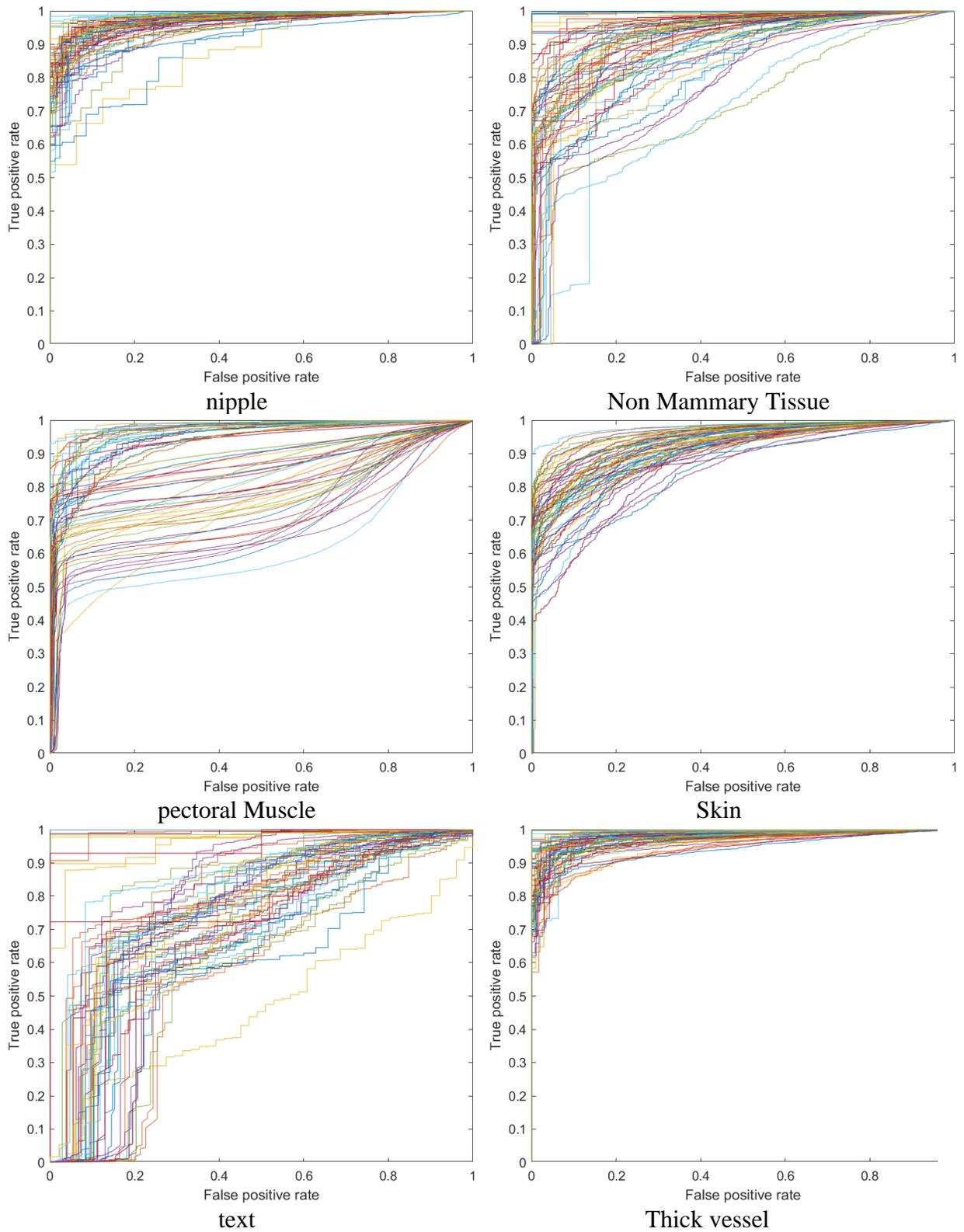

**Figure 10:** ROC cruves for of segmentation using UNet method with mamography images

**Table 2:** AUC value for detection of each class based on the ROC curve

| Regoin | background | axillary Lymph Nodes | calcifications | cancer | foreign Object | mammary Gland | nipple | non-Mammary Tissue | pectoral Muscle | skin | text | thick Vessels |
|---|---|---|---|---|---|---|---|---|---|---|---|---|
| AUC | 98.66% | 95.95% | 98.78% | 96.75% | 95.46% | 0.9416 | 96.69% | 91.20% | 85.84% | 92.46% | 76.57% | 97.49% |

Overall, our findings indicate that the U-Net method outperforms the pretrained SAM architecture in tumor detection for both BUS and mammographic images. The U-Net demonstrates superior performance in accurately predicting tumor areas, especially in challenging cases where the boundaries are not clearly defined or the tumor exhibits irregular shapes and margins. In contrast, SAM's performance is more limited, particularly in detecting malignant tumors and objects with weak boundaries or complex shapes.

The disparities in segmentation performance between benign and malignant tumors can be attributed to the inherent characteristics and complexities associated with these tumor types. Benign tumors often have well-defined borders, enabling precise segmentation. The U-Net successfully captures and defines the boundaries of such tumors, as they often exhibit a more regular and structured form. On the other hand, malignant tumors are characterized by higher levels of heterogeneity, irregular shapes, and margins due to their aggressive and invasive nature. These factors make accurate segmentation more challenging. As a result, the U-net's performance in identifying malignant tumors may experience a slight decrease compared to benign tumors. Moreover, the evaluation of the U-Net method on mammographic images revealed certain limitations. The presence of the pectoral muscle in the images posed a challenge in accurately identifying the tumor area, as the similarity in colors between the tumor and the muscle led to misclassification. Additionally, the U-Net had difficulty in detecting thick vessels in the breast tissue, often resulting in a limited number of points for their identification. These limitations highlight the need for further improvement and refinement of the U-Net architecture to address such challenges in mammographic image analysis.

### *4.4. Results of segmentation using SAM architecture with mammographic images*

In contrast, the pretrained SAM architecture demonstrated its ability to segment tumor tissue into different regions. However, it exhibited limitations in accurately identifying tumor areas within normal tissue due to color intensity differences. SAM's sensitivity to text and notes on the images also impacted its performance, as it falsely considered them as tumor areas. Furthermore, SAM's inability to identify regions for separation in some cases indicates its limitations in handling complex or ambiguous image content. The disparities between the two methods can be attributed to the underlying architectural differences and feature extraction capabilities. The U-Net architecture, specifically designed for medical image segmentation, leverages its deep learning framework and extensive training on the dataset to accurately identify tumor areas (see Figure 11).

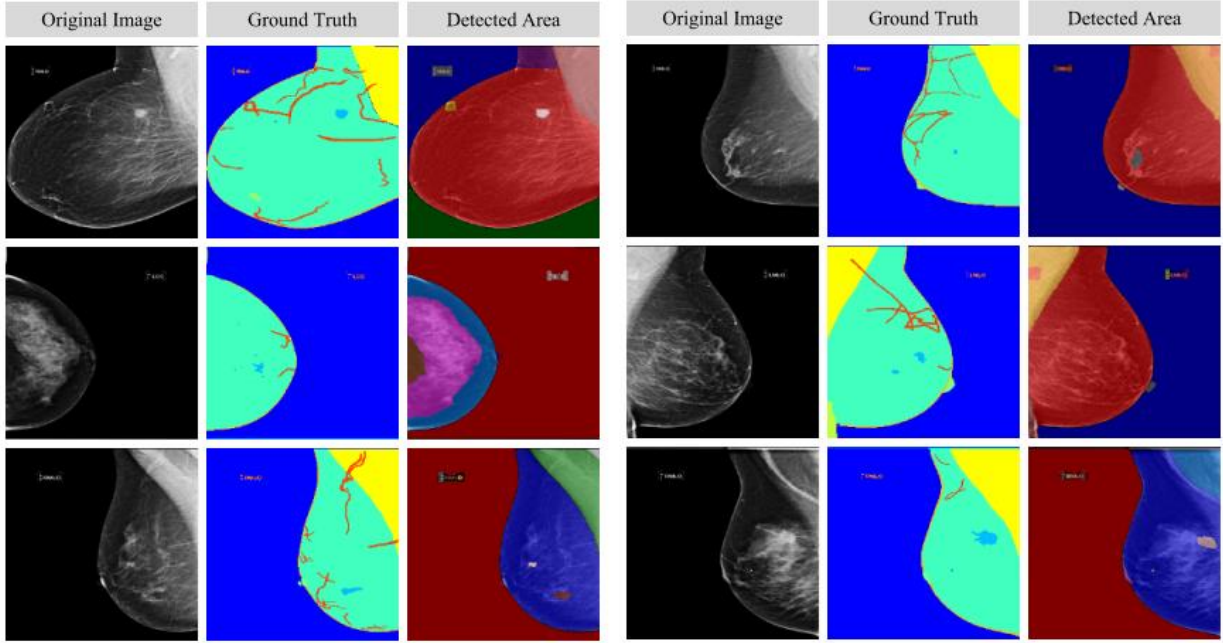

**Figure 11:** Results of segmentation using SAM method with mamography images

In contrast, the pretrained SAM architecture, although capable of segmenting certain medical objects effectively, relies heavily on human-provided cues or points for accurate segmentation. This dependency on explicit cues limits SAM's performance and generalizability across various tasks and datasets.

## 5. Discussion

To evaluate the performance of the proposed U-Net vs SAM method, we created a large-scale and diversified dataset containing 5000 breast tumor segmentation tasks in breast ultrasound images and 305 breast tumor segmentation tasks in mammography images. mammography imaging contains 13 predicted segmentation logits, including background, text, and 11 organ classes (axillary-lymph nodes, calcifications, cancer, foreign objects, mammary gland, nipple, non-mammary tissue, pectoral muscle, skin, thick vessels, unclassified).

### 5.1. Model Performance under different modalities

To validate the overall performance of our proposed U-Net model, we compared it to the SAM method using the two most commonly employed modalities for breast imaging, namely BUS and mammography, from the aforementioned breast segmentation datasets. We conducted a qualitative comparison of the segmentation masks obtained from SAM and U-net. The segmentation regions predicted by U-Net in both BUS and mammography images exhibited smoother contours and higher accuracy when compared to the results obtained from the SAM approach. Although the results indicate that SAM can accurately segment tumors in certain circumstances, a disparity still exists between the SAM and U-Net methods in the context of target medical image segmentation tasks.

SAM demonstrates good performance in segmenting benign tumors and other specified objects, particularly when applied to mammography. However, its performance was deemed inadequate for tumor detection in BUS due to the lower quality of ultrasound images it produces. Furthermore, SAM fell short when dealing with malignant tumors and other undefined structures. Conversely, U-Net achieved accurate segmentation even in challenging sections that are difficult to discern with the naked eye. It is worth noting that these findings underscore the necessity of fine-tuning a general segmentation model specifically for medical images to achieve optimal performance.

### 5.2. Factors potentially impacting SAM's accuracy

SAM's performance is easily influenced by numerous factors. The pre-trained SAM model performs well for segmenting organ regions with distinct borders, such as the nipple segmentation, large, connected objects, and benign tumors in BUS and particularly mammograms. Yet, may have failure to correctly identify dense object segmentation, such as malignant tumors, microcalcifications segmentation, and amorphous lesion areas. In both BUS and mammography imaging, SAM was unable to find some concealed items. Additionally, it is difficult to effectively segment targets with weak boundaries since the pre-trained SAM model is particularly prone to over-segmentation findings, especially in normal tissue of BUS images. Furthermore, the model might not pinpoint the proper segmentation target when the content of the imaging is heterogeneous. Likewise, SAM may produce outliers when segmenting targets with distinct boundaries if surrounding objects have good contrasts similarly. Overall, in both device evaluations, SAM's performance was noticeably worse than U-Net approaches. This is probably because the model's segmentation performance is strongly associated with the feature extraction capability of its backbone. Because of the complexities of medical imaging, the foreground and background are readily confused. Selecting some background points that are visually similar to the foreground ones will result in inaccurate guidance being generated for SAM. These results suggest that SAM has a poor perception of objects in medical imagery. Specifically, perception for objects with blurred boundaries or complex shapes. Our conclusion shows that, despite SAM's high perception of some medical objects, it frequently has difficulty correctly identifying and segmenting medical objects. Numerous authors noticed that SAM needs substantial human previous knowledge (i.e., points) in order to produce results that are reasonably accurate for various tasks. Otherwise, SAM produces incorrect segmentation, particularly when no cues are provided. The authors of [https://doi.org/10.48550/arXiv.2304.10517] used various point prompt counts to assess SAM's performance. They noticed that the SAM's performance converges as the number of points rises. SAM's performance is moderate overall and rather inconsistent across datasets and cases, they found. But to this point, U-Net has improved the model's capacity to identify challenging segmentation targets.

### 5.3. Benign vs malignant tumor detection with Different Modalities

This disparity in segmentation performance between cancerous and noncancerous breast tumors is likely due to differences in the characteristics and complexities of malignant and benign breast tumors. The following reasons may be the cause for segmentation performance discrepancies that have been observed. First off, benign tumors frequently have well-defined borders, which makes precise segmentation feasible. The model can accurately capture and define the boundaries of these tumors since they frequently display a more regular and structured form. Conversely, malignant breast tumors are characterized by higher levels of heterogeneity, irregular shapes, and margins. Their aggressive nature and invasive growth tendencies make precise segmentation complicated. Due to the increased complexity and variability of malignant tumors, it is more challenging to capture their entire extent, resulting in a small decrease in segmentation performance. Clearly, SAM performed better on mammograms than BUS due to the low image quality of the US dataset for these structures.

## 6. Conclusion

This study compared the performance of two tumor segmentation methods, U-Net and pretrained SAM, using breast ultrasound (BUS) and mammographic images for breast cancer detection. The evaluation was conducted on a diverse dataset comprising BUS images and mammographic images, encompassing both benign and malignant tumors. The results unequivocally demonstrated that the U-Net method outperformed the pretrained SAM architecture in accurately identifying tumor areas in both BUS and mammographic images. The U-Net exhibited superior performance in segmenting tumors with challenging characteristics, such as irregular shapes, indistinct boundaries, and high heterogeneity. Its ability to accurately capture and define tumor boundaries, even in difficult cases, highlights its robustness and effectiveness for tumor detection in breast imaging. In contrast, the pretrained SAM architecture exhibited limitations in accurately identifying tumor areas, especially in cases involving malignant tumors and objects with weak boundaries or complex shapes. SAM's performance was also influenced by the presence of text and notes on the images, which affected its segmentation results. These limitations indicate the need for further refinement and improvement of the SAM architecture to enhance its performance and generalizability in medical image segmentation tasks, particularly in the context of breast cancer detection. The disparities observed between the two methods can be attributed to their underlying architectural differences and feature extraction capabilities. The U-net, specifically designed for medical image segmentation, leverages its deep learning framework and extensive training on the dataset to accurately predict tumor areas. In contrast, SAM heavily relies on human-provided cues or points for accurate segmentation, making it more dependent on explicit guidance and less adaptable to various tasks and datasets.

Furthermore, our evaluation revealed differences in segmentation performance between benign and malignant tumors. Benign tumors, characterized by well-defined borders and regular structures, were more accurately segmented by both methods. However, the segmentation of malignant tumors, with higher heterogeneity, irregular shapes, and margins, posed greater challenges for both U-Net and SAM. Nevertheless, the U-Net demonstrated a slight advantage in identifying malignant tumors compared to SAM, showcasing its potential for accurate detection even in complex tumor cases. In the evaluation of mammographic images, the U-Net showcased its strengths in tumor detection, except in cases where the presence of the pectoral muscle led to misclassification of tumor areas. Additionally, the U-Net faced challenges in identifying thin vessels within the breast tissue, resulting in limited points for their detection. These limitations highlight the need for further advancements and adaptations in the U-Net architecture to overcome these challenges and improve its performance in mammographic image analysis, particularly for cases involving the pectoral muscle and thin vessels. The findings of this study emphasize the importance of selecting appropriate segmentation models, specifically tailored for medical images, to achieve optimal performance in breast cancer detection. The U-Net method demonstrated its potential as a robust and accurate tumor detection tool, particularly in challenging cases. However, further research is required to enhance its performance and address the identified limitations, such as the misclassification of tumor areas in the presence of certain anatomical structures. In conclusion, this study contributes to the field of medical image segmentation by comparing the performance of U-Net and pretrained SAM methods for tumor detection in breast cancer. The results highlight the superiority of the U-Net architecture, while also identifying areas for improvement. These findings can guide future research and development efforts in the field of tumor detection, ultimately enhancing the accuracy and efficiency of breast cancer diagnosis and treatment planning.

Future work in this field could focus on several areas to further improve tumor detection and segmentation in breast cancer. Firstly, exploring advanced deep learning techniques and architectures, such as attention mechanisms or hybrid models, could enhance the performance of existing methods like U-net. Additionally, incorporating multimodal imaging data, such as combining ultrasound and mammography images, could provide more comprehensive information for accurate tumor segmentation. Furthermore, investigating the use of transfer learning and domain adaptation techniques to improve the generalizability of segmentation models across different datasets and clinical sites would be beneficial. Finally, conducting extensive validation studies and clinical trials to assess the real-world impact and effectiveness of these segmentation methods is essential for their successful translation into clinical practice.